\documentclass[aps,pre,twocolumn]{revtex4-1}

\usepackage{amsmath}
\usepackage{amssymb}
\usepackage{graphicx}
\usepackage{color}
\usepackage{bm}
\usepackage{subfigure}
\usepackage{multirow}
\usepackage{latexsym}
\usepackage{sidecap}
\usepackage{comment}

\begin{document}
	
\title{Self-assembly of polyhedral bilayer vesicles from Piezo ion channels}

\author{Mingyuan Ma}

\author{Christoph A. Haselwandter}

\affiliation{Department of Physics and Astronomy and Department of Quantitative and Computational Biology, University of Southern California, Los Angeles, CA 90089, USA}


\begin{abstract}

Piezo ion channels underlie many forms of mechanosensation in vertebrates, and have been found 
to bend the membrane into strongly curved dome shapes. We develop here a methodology describing the self-assembly of lipids and Piezo into polyhedral bilayer vesicles. We validate this methodology for bilayer vesicles formed from bacterial mechanosensitive channels of small conductance, for which experiments found a polyhedral arrangement of proteins with snub cube symmetry and a well-defined characteristic vesicle size. On this basis, we calculate the self-assembly diagram for polyhedral bilayer vesicles formed from Piezo. We find that the radius of curvature of the Piezo dome provides a critical control parameter for the self-assembly of Piezo vesicles, with high abundances of Piezo vesicles with octahedral, icosahedral, and snub cube symmetry with increasing Piezo dome radius of curvature.

\end{abstract}

\maketitle

The ability to sense mechanical stimuli such as touch and changes in fluid pressure is fundamental to life. Piezo ion channels \cite{coste10} have recently been found to provide the molecular basis for a wide range of seemingly unrelated forms of mechanosensation in vertebrates \cite{ranade15,honore15,murthy17,wu17,gottlieb17,parpaite17,douguet19,dance20}. Structural studies \cite{Guo2017,saotome18,zhao18,wang19} have demonstrated that Piezo is an unusually large ion channel that locally bends the membrane into the approximate shape of a spherical dome. Only closed-state structures of Piezo, obtained in the absence of transmembrane gradients, are currently available. The interaction of Piezo with the surrounding lipid membrane has been investigated through electron microscopy experiments in which Piezo proteins were embedded in lipid bilayer vesicles \cite{Guo2017,lin19}. The highly curved shape of the (closed-state) Piezo dome yields pronounced shape deformations in the surrounding lipid membrane \cite{Guo2017,Haselwandter2018,lin19}. These shape deformations may play an important role in Piezo gating \cite{Guo2017,saotome18,zhao18,wang19,Haselwandter2018,lin19}, with transitions from closed to open states of Piezo potentially being accompanied by an increase in the radius of curvature of the Piezo dome.

Electron cryotomography experiments on membrane protein polyhedral nanoparticles (MPPNs) formed from bacterial mechanosensitive channels of small conductance (MscS) \cite{Wu2013,basta14,Bass2002,Steinbacher2007} have shown that lipids and membrane proteins can self-assemble into lipid bilayer vesicles with a polyhedral protein arrangement and a well-defined characteristic size, thus facilitating structural studies. Similarly, MPPNs formed from Piezo may permit structural analysis of Piezo in the presence of transmembrane gradients \cite{basta14,zhang03,liu04,cockburn04} and aid the further investigation of the interaction of Piezo with the surrounding lipid membrane. In this Letter we develop a model of the self-assembly of MPPNs from Piezo ion channels. Our theoretical approach is based on a previous mean-field model \cite{li16,li17} in which the observed symmetry and size of MPPNs formed from MscS were found to emerge from the interplay of protein-induced lipid membrane deformations, topological defects in protein packing in MPPNs, and thermal effects. This previous model relied on the assumption that membrane proteins in MPPNs only weakly curve the membrane, which is a suitable assumption for MscS \cite{Bass2002,Steinbacher2007,Phillips2009} but not Piezo \cite{Guo2017,saotome18,zhao18,wang19,lin19,Haselwandter2018}. We first develop a general methodology for predicting the symmetry and size of MPPNs composed of proteins that may induce arbitrarily large membrane shape deformations. We validate this methodology for MPPNs formed from MscS. On this basis, we then calculate the self-assembly diagram for MPPNs formed from Piezo. We find that the radius of curvature of the Piezo dome provides a critical control parameter for the self-assembly of MPPNs from Piezo proteins, with high abundances of MPPNs with octahedral, icosahedral, and snub cube symmetry as the radius of curvature of the Piezo dome is increased. Our analysis suggests that, under suitable conditions, self-assembly of MPPNs from Piezo proteins results in highly symmetric MPPNs with a well-defined characteristic size.

\begin{figure}[t!]
	\centering
	\includegraphics[width=\columnwidth]{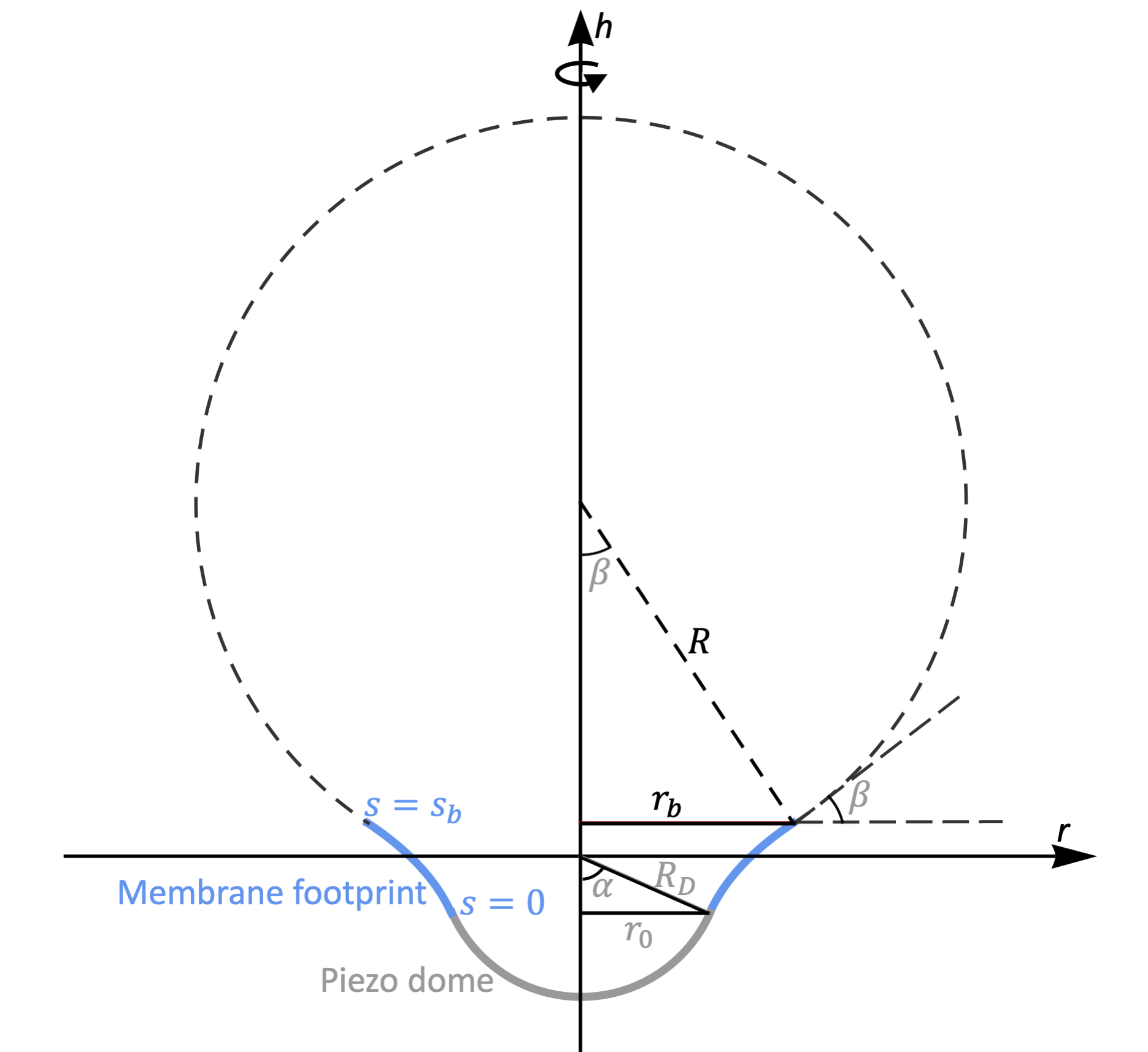} 
	\caption{Schematic of MPPNs formed from Piezo ion channels. The thick gray curve shows the Piezo dome with radius of curvature $R_D$ and cap angle $\alpha$. The blue curve shows the membrane footprint of the Piezo dome with arclengths $s=0$ and $s=s_b$ at the inner and outer membrane footprint boundaries, respectively. At $s=s_b$, the Piezo membrane footprint is assumed to connect smoothly to the membrane footprints associated with the neighboring Piezo proteins on the MPPN surface, with contact angle $\beta$. We denote the inner and outer membrane footprint boundaries along the $r$-axis by $r_0$ and $r_b$, respectively. The MPPN radius is given by $R = r_b/\sin\beta$. For each $s_b$, Piezo's membrane footprint is completely determined by a given set of values of $\alpha$, $R_D$, $r_0=R_D \sin\alpha$, and $\beta$, which are indicated in gray. For simplicity, we only show here one of the $n$ proteins on the MPPN surface.
	}
	\label{Fig1}
\end{figure}

\textit{Statistical thermodynamics of MPPN self-assembly.} As detailed in Refs.~\cite{li16,li17,Ma20}, the formalism describing the statistical thermodynamics of micelle and viral capsid self-assembly \cite{BenShaul1994,Safran2003,Bruinsma2003} successfully predicts the observed symmetry and size of MPPNs formed from MscS \cite{Wu2013,basta14}. We adapt here this formalism to explore the self-assembly of MPPNs formed from Piezo ion channels. In particular, we take MPPNs to be in the thermodynamic equilibrium state minimizing the Helmholtz free energy $F = U-T S$, where $U$ is the internal energy of the system and $T$ and $S$ are the entropy and temperature of the system, respectively. MPPNs formed from MscS were obtained experimentally \cite{basta14,Wu2013} in dilute aqueous solutions with a protein number fraction $c\approx 7.8\times10^{-8}\ll 1$,~where
\begin{equation}\label{eqcExp}
c = \sum_{n}\frac{N_n}{N_w}\,,
\end{equation}
in which $N_n$ denotes the total number of proteins bound in MPPNs with $n$ proteins each and $N_w$ denotes the total number of solvent (water) molecules in the system. In this dilute limit with no interactions between MPPNs, $S$ is given by the mixing entropy \cite{BenShaul1994,Safran2003}
\begin{equation}\label{eqS}
S = -N_w k_B\sum_{n}\Phi(n)[\ln \Phi(n) - 1]\,,
\end{equation}
where $k_B$ is Boltzmann's constant and the MPPN number fraction $\Phi(n) = N_n/n N_w$. Similarly, we have the MPPN internal energy \cite{BenShaul1994,Safran2003,li16,li17,Ma20}
\begin{equation}\label{eqU}
U = N_w\sum_{n}\Phi(n)E_\text{min}(n)\,,
\end{equation}
where $E_\text{min}(n)$ is the minimum energy of MPPNs with $n$ proteins each. We obtain $E_\text{min}(n)$ by minimizing the MPPN energy $E(n,R)$, at each $n$, with respect to the MPPN radius at the bilayer midplane, $R$ (see Fig.~\ref{Fig1}). The dominant symmetry and size of MPPNs observed in experiments on MPPNs formed from MscS are successfully predicted by considering contributions to $E$ due to membrane bending deformations, $E_b$, and topological defects in protein packing in MPPNs, $E_d$, such that $E=E_b + E_d$ \cite{Wu2013,basta14,li16,li17}. We show below how $E_\text{min}(n)$ can be calculated from $E_b$ and $E_d$ for MPPNs with the large membrane curvatures induced by the Piezo dome \cite{Guo2017,Haselwandter2018,lin19}.

Minimization of $F$ with respect to $\Phi(n)$ results in \cite{li16,li17,BenShaul1994,Safran2003,Bruinsma2003}
\begin{equation}\label{defPhi}
\Phi(n)=e^{[\mu n-E_{\text{min}}(n)]/k_B T}\ ,
\end{equation}
where the protein chemical potential $\mu$ is determined by the constraint 
\begin{equation} \label{consmu}
\sum_n n \Phi(n)=c
\end{equation}
imposing the fixed protein number fraction in Eq.~(\ref{eqcExp}). In our previous work on MPPN self-assembly \cite{li16,li17,Ma20} we restricted $n$ to the range $10 \leq n \leq 80$. As shown below, the strongly curved shape of the Piezo dome means that the self-assembly diagram for MPPNs formed from Piezo can be dominated by MPPN states with $n<10$. We allow here for the $n$-range $3 \leq n \leq 80$, with the smallest number of proteins per MPPN permitting the formation of a polyhedral structure corresponding to $n=4$. The MPPN equilibrium distribution is then given by
\begin{equation} \label{eqdefphi}
\phi(n)=\frac{\Phi(n)}{\sum_{n=3}^{80} \Phi(n)}\,,
\end{equation}
where $\phi(n)$ is the fraction of MPPNs containing $n$ proteins each and $\Phi(n)$ is obtained from Eq.~(\ref{defPhi}) with Eqs.~(\ref{eqcExp}) and~(\ref{consmu}).

\textit{Nonlinear MPPN shape equations.} We describe here the Piezo dome as a spherical cap with area $S_\text{cap} \approx 390~\mathrm{nm}^2$ and radius of curvature $R_D$ \cite{Guo2017,saotome18,zhao18,wang19} (Fig.~\ref{Fig1}). Assuming that MPPNs are under negligible membrane tension \cite{li17}, the shape and energy of Piezo's membrane footprint can be estimated \cite{Haselwandter2018} by minimizing the bending energy of the lipid membrane \cite{Canham1970,Helfrich1973,Evans1974},
\begin{equation}\label{eqdefG}
G_b = \frac{K_b}{2} \int dA \left(c_1 + c_2\right)^2 \,,
\end{equation}
where $K_b$ is the lipid bilayer bending rigidity, $c_{1}$ and $c_{2}$ are the local principal curvatures of the mid-bilayer surface, and the integral runs over Piezo's membrane footprint. Previous experiments on MPPNs formed from MscS employed lipids with $K_b\approx14$~$k_B T$ \cite{basta14,Wu2013,Rawicz2000}. We use this value of $K_b$ throughout this~Letter.

Membrane-mediated interactions between Piezo proteins are expected to favor approximately hexagonal protein arrangements
\cite{Gozdz2001,Auth2009,Muller2010,Fournier1999,Weitz2013}. The bending energy of MPPNs can then be estimated from a mean-field approach \cite{Gozdz2001,Auth2009,Muller2010,li16,li17,Ma20} in which the boundary of the hexagonal unit cell of the protein lattice is approximated by a circle. In particular, we divide the surface of MPPNs containing $n$ Piezo proteins into $n$ identical, circular membrane patches, each with a Piezo dome at its center (Fig.~\ref{Fig1}). Using the arclength parameterization of surfaces, Eq.~(\ref{eqdefG}) can be rewritten as \cite{Peterson1985,Seifert1991,Seifert1994,kuhnel15}
\begin{eqnarray}\nonumber
G_b &=& \int_{0}^{s_b} ds \bigg[\pi K_b r \left(\dot{\psi} + \frac{\sin\psi}{r}\right)^2 \\&&+ \lambda_r(s) (\dot{r} - \cos\psi) + \lambda_h(s) (\dot{h}-\sin\psi)\bigg]  \label{eqArcG}
\end{eqnarray}
for each membrane patch, where $s$ is the arclength along the profile of Piezo's membrane footprint, $s=0$ at the inner boundary of Piezo's membrane footprint (the boundary of the Piezo dome) and $s=s_b$ at the outer boundary of Piezo's membrane footprint away from the Piezo dome, $h(s)$ denotes the height of Piezo's membrane footprint along its symmetry axis $h$, $r(s)$ denotes the radial coordinate of Piezo's membrane footprint perpendicular to the $h$-axis, $\psi(s)$ denotes the angle between the tangent to the profile of Piezo's membrane footprint and the $r$-axis, and the Lagrange parameter functions $\lambda_r(s)$ and $\lambda_h(s)$ enforce the geometric constraints $\dot{r} = \cos\psi$ and $\dot{h}=\sin\psi$ inherent in the arclength parameterization of surfaces (Fig.~\ref{Fig1}).

The boundary conditions on Piezo's membrane footprint at the Piezo dome boundary follow from the assumption that the membrane surface is smooth at $s=0$ \cite{Guo2017,Haselwandter2018} (Fig.~\ref{Fig1}):
\begin{eqnarray}\label{eqbr0}
r(0) &\equiv& r_0 = R_D \sin \alpha \,,\\ \label{eqbh0}
h(0) &=& -R_D \cos \alpha\,,\\
\psi(0) &=& \alpha\,,
\end{eqnarray}
with the membrane-Piezo dome contact angle \cite{weisstein17}
\begin{equation}
\alpha= \cos^{-1}\left(1-\frac{S_\text{cap}}{2 \pi R_D^2}\right)\,.
\end{equation}
Denoting the contact angle at the outer boundary of Piezo's membrane footprint by $\beta$ and the solid angle associated with each unit cell on the MPPN surface by $\Omega$, we have $\Omega = 2\pi(1-\cos\beta)$. Since each unit cell on the MPPN surface contains one protein, we also have $\Omega = 4\pi/n$, resulting in the boundary condition \cite{Auth2009,li16,li17,Ma20}
\begin{equation}\label{eqbpsib}
\psi(s_b) = \cos^{-1}\left(1-\frac{2}{n}\right)
\end{equation}
at $s=s_b$ (Fig.~\ref{Fig1}). Equations~(\ref{eqbr0})--(\ref{eqbpsib}) encapsulate the effects of a particular Piezo dome shape and protein number per MPPN on $E_b$. With the (arbitrary) origin of the $r$-$h$ coordinate system fixed via Eqs.~(\ref{eqbr0}) and~(\ref{eqbh0}) (Fig.~\ref{Fig1}), we assume that the values of $r(s_b) \equiv r_b$ and $h(s_b)$ can be freely varied when finding the extremal functions of Eq.~(\ref{eqArcG}) \cite{Hilbert1953,vanbrunt04}, resulting in the natural boundary conditions
\begin{eqnarray}
\label{eqbprb}
p_r(s_b) &=& \lambda_r(s_b) = 0\,,\\
\label{eqbphb}
p_h(s_b) &=& \lambda_h(s_b) = 0
\end{eqnarray}
at $s=s_b$, where $p_r(s)\equiv \partial L/\partial \dot{r}$ and $p_h(s)\equiv \partial L/\partial \dot{h}$ are the generalized momenta associated with the generalized displacements $r(s)$ and $h(s)$, in which the Lagrangian function $L$ is given by the integrand in Eq.~(\ref{eqArcG}).

To determine the stationary shapes of MPPNs we solve the Hamilton equations associated with the membrane bending energy in Eq.~(\ref{eqArcG}) \cite{Kibble2004,Deserno2003,Deserno2004,Nowak2008,Zhang2008,Hashemi2014,Foret2014,Haselwandter2018} subject to the boundary conditions on Piezo's membrane footprint in Eqs.~(\ref{eqbr0})--(\ref{eqbphb}). Compared to the corresponding Euler-Lagrange equations associated with Eq.~(\ref{eqArcG}) \cite{Hilbert1953,vanbrunt04,Kibble2004,Peterson1985,Seifert1991,Seifert1994,agudo16,bahrami16} these Hamilton equations are of first rather than second order in derivatives. The Hamilton equations for Eq.~(\ref{eqArcG}) are given by
\begin{eqnarray} \label{eqHamilton0}
\dot{\psi} &=& \frac{p_\psi}{2r} - \frac{\sin\psi}{r}\,,\\
\dot{r} &=& \cos\psi\,,\\
\dot{h} &=& \sin\psi\,,\\
\dot{p}_\psi &=& \left(\frac{p_\psi}{{r}} - p_h\right)\cos\psi + p_r\sin\psi\,,\\
\dot{p_r} &=& \frac{p_\psi}{r} \left(\frac{p_\psi}{4r} - \frac{\sin\psi}{r} \right)\,, \label{eqHamilton1}\\
\dot{p_h} &=& 0\,,
\label{eqHamilton12}
\end{eqnarray}
where $p_\psi(s)\equiv \partial L/\partial \dot{\psi}$ is the generalized momentum associated with the generalized displacement $\psi(s)$. The boundary condition in Eq.~(\ref{eqbphb}) and the Hamilton equation in Eq.~(\ref{eqHamilton12}) imply that $p_h(s)=0$ for $0 \leq s \leq s_b$. The solutions of the remaining five Hamilton equations in Eqs.~(\ref{eqHamilton0})--(\ref{eqHamilton1}) are specified by the five boundary conditions in Eqs.~(\ref{eqbr0})--(\ref{eqbprb}). A numerical difficulty arises here in that some of these boundary conditions are specified at $s=0$, while others are specified at $s=s_b$. We thus solve Eqs.~(\ref{eqHamilton0})--(\ref{eqHamilton1}) using a shooting method \cite{Peterson1985,Seifert1991,Seifert1994,agudo16,bahrami16,Deserno2003,Deserno2004,Nowak2008,Zhang2008,Hashemi2014,Foret2014,Haselwandter2018,burden11,gautschi12}, for which we introduce the boundary conditions
\begin{eqnarray}
p_{\psi}(0) &=& p_{\psi;0}\,, \\
p_{r}(0) &=& p_{r;0}\,,
\end{eqnarray}
where $p_{\psi;0}$ and $p_{r;0}$ must be adjusted so as to satisfy the boundary conditions in Eqs.~(\ref{eqbpsib}) and~(\ref{eqbprb}). The values of $p_{\psi;0}$ and $p_{r;0}$ can be conveniently determined through the \textit{FindRoot} command in \textit{Mathematica} \cite{mathematica17}. We obtain $E_b$ by substituting the solutions of Eq.~(\ref{eqHamilton0})--(\ref{eqHamilton12}) into Eq.~(\ref{eqArcG}) and (numerically) integrating with respect to $s$.

In the above numerical calculation of $E_b$, the size of Piezo's membrane footprint enters through the value of $s_b$. We note that the MPPN radius $R$ is related to the in-plane membrane patch radius $r_b$ via $R = r_b/\sin\beta$ (Fig.~\ref{Fig1}). In general, different values of $s_b$ yield different values of $r_b$ and, hence, $R$ for the stationary membrane footprints. We therefore minimize, at each $n$, $E_b$ with respect to the MPPN size by adjusting $s_b$ so that $E_b$ is minimal, and then determine the values of $r_b$ and $R$ associated with this value of $s_b$. We used here the $s_b$-range $0.01$~$\text{nm}\leq s_b \leq 20$~nm, and generally employed a resolution $\Delta s_b=0.2$~nm for our numerical calculations. However, we found that, within the range $0.01$~$\text{nm}\leq s_b \leq 2$~nm, $E_b$ may vary rapidly with $s_b$, and therefore used a finer resolution $\Delta s_b=0.05$~nm within this range of $s_b$. We interpolated $E_b$ between the values of $s_b$ considered here using third-order splines. For each Piezo dome shape and protein number per MPPN considered, the value of $s_b$ minimizing $E_b$ thus specifies the MPPN size with minimal bending energy.

\textit{Small-gradient approximation for MPPNs.} In the Monge parameterization of Eq.~(\ref{eqdefG}), $h$ is regarded as a single-valued function of $r$, $h(r)$ \cite{kuhnel15}. In the small-gradient limit of the Monge parameterization, $|\nabla h| \ll 1$, the stationary membrane shapes implied by Eq.~(\ref{eqdefG}) can be solved for analytically \cite{Auth2009}, yielding an exact expression for the membrane bending energy in MPPNs. In particular, denoting the $E_b$ obtained for $|\nabla h| \ll 1$ by $\bar E_b$, we have~\cite{Auth2009,li16,li17,Ma20}
\begin{equation} \label{eq:Eh}
\bar E_b(n,R)=\frac{{2n \pi K_b \left(r_0 \tan \alpha -R \sin \beta \tan \beta \right)}^2}{R^2 \sin^2 \beta-r_0^2}\,.
\end{equation}
Equation~(\ref{eq:Eh}) was employed previously \cite{li16,li17,Ma20} to construct the MPPN self-assembly diagram for MPPNs formed from MscS proteins, which only weakly curve the membrane. For the purposes of this Letter, the analytic solution in Eq.~(\ref{eq:Eh}) presents a useful reference point for the fully nonlinear, numerical solutions obtained from Eq.~(\ref{eqArcG}).

\textit{Topological defects in protein packing.} The spherical shape of MPPNs necessitates topological defects in the preferred hexagonal packing of proteins, which incur an $n$-dependent energy penalty. At the mean-field level, deviations from hexagonal protein packing due to the spherical shape of MPPNs can be quantified for a given $n$, in analogy to viral capsids \cite{Bruinsma2003}, through the fraction of the surface of a sphere enclosed by $n$ identical non-overlapping circles at closest packing \cite{li16,li17,Ma20}, $p(n)$. We use here the values of $p(n)$, and associated symmetries of protein packing in MPPNs, compiled in Refs.~\cite{Clare1986,Clare1991}. Approximating the spring network associated with the energetically preferred hexagonal protein arrangement by a uniform elastic sheet, the leading-order contribution to the MPPN defect energy is thus given by \cite{Bruinsma2003,li16,li17,Ma20}
\begin{equation}\label{eq:Ed}
E_d(n,R)=\frac{K_s}{2} A \left[\frac{p_\mathrm{max}-p(n)}{p_\mathrm{max}}\right]^2,
\end{equation}
where $K_s$ is the stretching modulus of the elastic sheet, $A$ is the MPPN surface area, and $p_{\text{max}}=\pi/2\sqrt{3}$ denotes the optimal packing fraction associated with a uniform hexagonal protein arrangement. As detailed in Refs.~\cite{li16,li17,Ma20}, the stretching modulus in Eq.~(\ref{eq:Ed}) is given by 
\begin{equation} \label{eq:Ks}
K_s = \frac{\sqrt{3}}{24 n}\frac{\partial^2E_b}{\partial r_b^2} \bigg|_{r_b=r_b^\mathrm{min}}\,,
\end{equation} 
where $r_b^\mathrm{min}$ is the value of $r_b$ that yields, for a given $n$, the minimal $E_b$.

Most straightforwardly, the MPPN surface area $A$ in Eq.~(\ref{eq:Ed}) can be approximated via $A=A_S$ \cite{li16,li17,Ma20}, where $A_S=4\pi R^2$ is the surface area associated with a spherical MPPN shape. The approximation $A=A_S$ breaks down for large enough deviations from a spherical shape, which is expected to be the case for MPPNs formed from Piezo. As an alternative to $A=A_S$, we therefore consider here the choice $A=A_D$ with the area of the deformed MPPN surface, $A_D$, being given by
\begin{equation} \label{AF}
A_D = n S_\text{cap} + 2 \pi n \int_{0}^{s_b} ds r \,,
\end{equation}
where, as noted above, $S_\text{cap}=390$~nm$^2$ for the Piezo dome. For MPPNs formed from MscS proteins, we approximate the MPPN surface area occupied by MscS by the spherical cap area $S_\text{cap} = 2\pi R^2 \left(1-\cos [\arcsin(r_0/R)] \right)$ in Eq.~(\ref{AF}), in which the MscS in-plane radius $r_0\approx 3.2$~nm \cite{li16,Bass2002,Steinbacher2007}. Unless indicated otherwise, we use here $A=A_D$ when evaluating Eq.~(\ref{eq:Ed}).

\textit{Constructing MPPN self-assembly diagrams.} We construct MPPN self-assembly diagrams from the fraction of MPPNs containing $n$ proteins each, $\phi(n)$ in Eq.~(\ref{eqdefphi}). To this end, we obtain the minimal MPPN energy, $E_\text{min}$ in Eq.~(\ref{eqU}), at each $n$ by minimizing the sum of $E_b$, calculated from the stationary $G_b$ implied by the arclength or Monge parameterization of Eq.~(\ref{eqdefG}), and the corresponding $E_d$ in Eq.~(\ref{eq:Ed}) with respect to the MPPN radius $R$. We perform this minimization subject to steric constraints due to the finite size of lipids and proteins. To determine the resulting constraints on $R$ we note that, at closest packing, the area of a sphere of radius $R$ enclosed by $n$ non-overlapping circles is given by $4\pi R^2 p(n)$ \cite{Clare1986,Clare1991}. Thus, each circular membrane patch occupies an area $4\pi R^2 p(n)/n$ and, as a result, suspends an angle $\bar \beta =  \cos^{-1} \left[1 - 2 p(n)/n\right]$ with respect to the MPPN center. Following Refs.~\cite{li16,li17,Ma20} we assume that the MPPN size must be large enough so that the Piezo dome at the center of each membrane patch is surrounded by at least one layer of lipids, resulting in the steric constraint 
\begin{equation}
R \geq \frac{r_0 + \rho_l}{\sin\bar{\beta}} \,,
\end{equation}
where $\rho_l$ denotes the lipid radius. For the lipids employed in experiments on MPPNs formed from MscS \cite{basta14,Wu2013} we have $\rho_l\approx0.45$~nm \cite{Damodaran1993}, which we use throughout this Letter. Once $E_\text{min}(n)$ is calculated, the $\phi(n)$ are conveniently obtained via Eq.~(\ref{consmu}) for arbitrary protein number fractions $c$. In the arclength parameterization of Eq.~(\ref{eqdefG}), changes in $R_D$ or $\alpha$ necessitate repeated numerical solution of Eqs.~(\ref{eqHamilton0})--(\ref{eqHamilton1}). We determine $\phi(n)$ for a discrete set of values of $R_D$ or $\alpha$ and interpolate $\phi(n)$ between these values, using third-order splines, to find the dominant $n$-states of MPPNs showing the largest values of $\phi(n)$ (see below).

\begin{figure}[t!]
	\centering
	\includegraphics[width=\columnwidth]{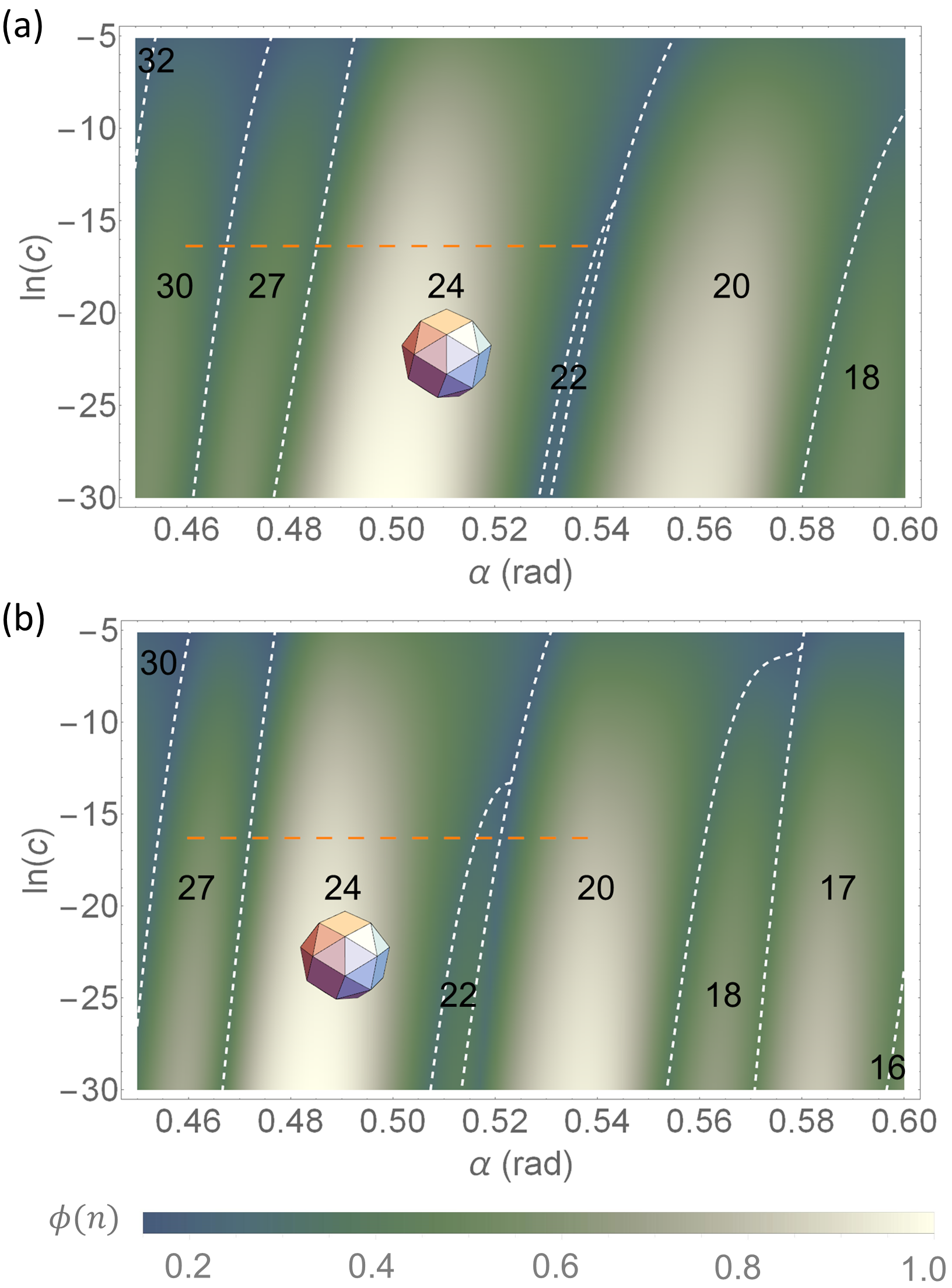} 
	\caption{MPPN self-assembly diagrams for MPPNs formed from MscS obtained using (a) the arclength parameterization of Eq.~(\ref{eqdefG}) and (b) the Monge parameterization of Eq.~(\ref{eqdefG}). The horizontal axes show the bilayer-protein contact angle, $\alpha$, and the vertical axes show the protein number fraction in solution, $c$. The dominant $n$-states of MPPNs are indicated by integers. The white dashed curves show transitions in the dominant MPPN $n$-states, with the colors indicating the maximum $\phi(n)$ among all MPPN $n$-states considered. We use the same color bar in panels (a) and (b). The dashed horizontal lines indicate the parameter values $c\approx 7.8\times10^{-8}$ and $\alpha\approx0.46$--$0.54$~rad corresponding to experiments on MPPNs formed from MscS \cite{Wu2013,basta14}, in which the snub cube with $n=24$ MscS proteins was found to provide the dominant MPPN symmetry (models of the snub cube below $n=24$).
	}
	\label{Fig2}
\end{figure}

\begin{figure}[t!]
	\centering
	\includegraphics[width=\columnwidth]{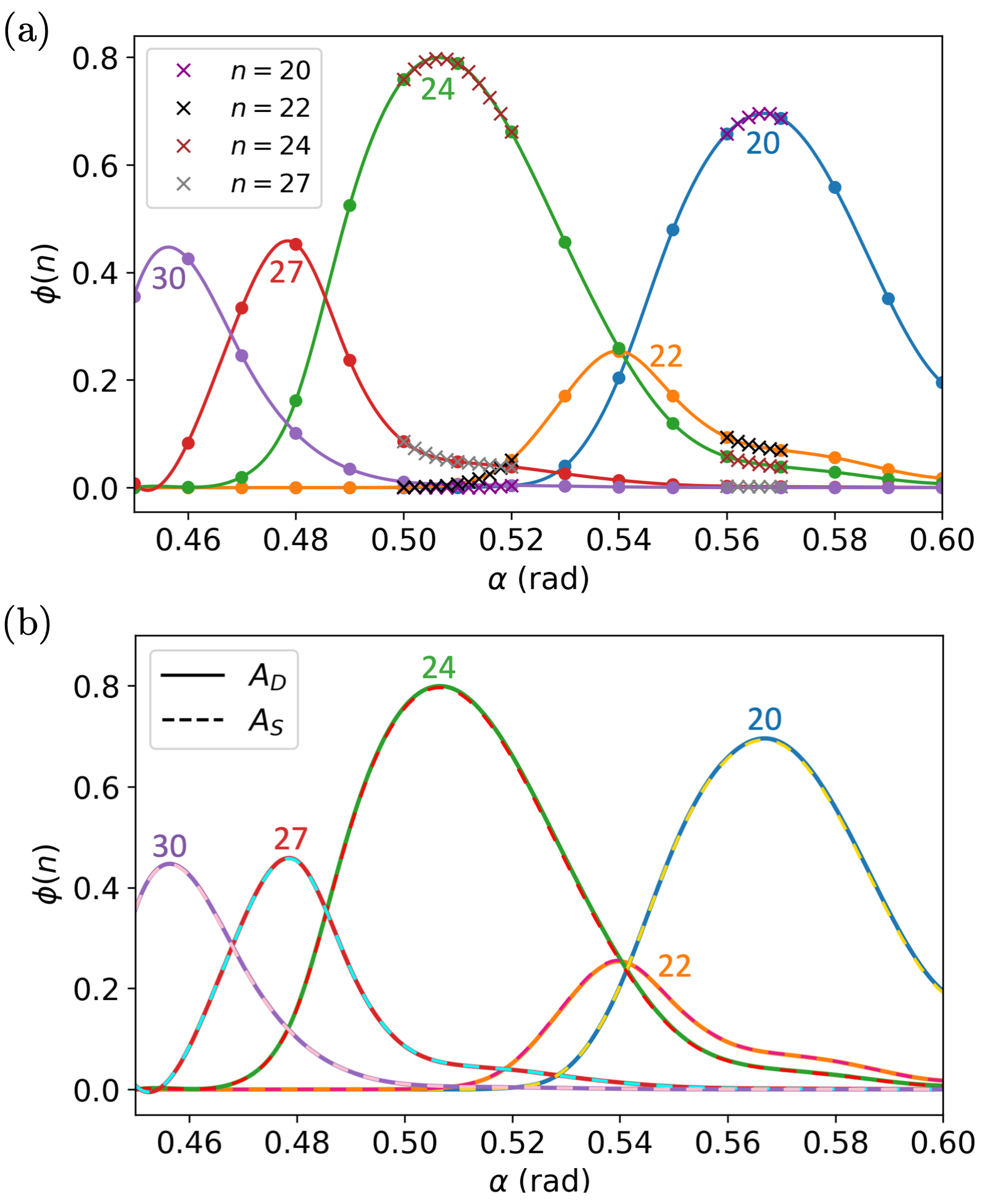} 
	\caption{Fractional abundance of MPPN $n$-states, $\phi(n)$, versus bilayer-protein contact angle, $\alpha$, for selected (dominant) MPPN $n$-states in Fig.~\ref{Fig2}(a) (indicated by integers) obtained with the arclength parameterization of Eq.~(\ref{eqdefG}) and the protein number fraction $c\approx 7.8\times10^{-8}$ used for the MPPN self-assembly experiments in Refs.~\cite{Wu2013,basta14}. As in Fig.~\ref{Fig2}(a), we calculated all curves by interpolation with respect to $\alpha$ of numerical results at a resolution $\Delta \alpha = 0.01$~rad [dots in panel (a)], using third-order splines. Panel (a) compares these interpolations with the corresponding results obtained at a finer resolution $\Delta \alpha=0.002$~rad (crosses). Panel~(b) compares the curves in panel (a) obtained using $A=A_D$ in Eq.~(\ref{AF}) (solid curves) with the corresponding results obtained using $A=A_S$ (dashed curves).
	}
	\label{Fig3}
\end{figure}

\textit{MPPN self-assembly from MscS.} Since MscS proteins only weakly curve the membrane \cite{Bass2002,Steinbacher2007,Phillips2009}, the Monge parameterization of Eq.~(\ref{eqdefG}) is expected to yield a good approximation of the self-assembly diagram of MPPNs formed from MscS. Figures~\ref{Fig2}(a) and~\ref{Fig2}(b) show the self-assembly diagrams for MPPNs formed from MscS obtained from the arclength and Monge parameterizations of Eq.~(\ref{eqdefG}), respectively. Both parameterizations of Eq.~(\ref{eqdefG}) predict, with no free parameters, that the snub cube ($O$-symmetry; $n=24$) provides the dominant MPPN symmetry for the $\alpha$-range $\alpha\approx0.46$--$0.54$~rad associated with MscS \cite{Bass2002,Steinbacher2007,Phillips2009,li16} and the protein number fraction $c \approx  7.8\times10^{-8}$ used in experiments on MPPNs formed from MscS \cite{Wu2013,basta14} (dashed horizontal lines in Fig.~\ref{Fig2}). Furthermore, the arclength and Monge parameterizations of Eq.~(\ref{eqdefG}) predict that the dominant MPPNs with snub cube symmetry have a characteristic bilayer midplane radius $9.8~\mathrm{nm} \lessapprox R \lessapprox 10$~nm and $9.8~\mathrm{nm} \lessapprox R \lessapprox 11$~nm, respectively. These predictions of Eq.~(\ref{eqdefphi}) are in quantitative agreement with experiments on MPPNs formed from MscS \cite{Wu2013,basta14}. A notable discrepancy between the MPPN self-assembly diagrams predicted by the arclength and Monge parameterizations of Eq.~(\ref{eqdefG}) is that, for the Monge parameterization, a given dominant MPPN $n$-state tends to appear at slightly smaller values of $\alpha$. For the $\alpha$-range relevant for MscS, this means that subdominant MPPNs tend to have larger $n$ in Fig.~\ref{Fig2}(a) than in Fig.~\ref{Fig2}(b). Since the Monge parameterization becomes less accurate as $\alpha$ is increased, these shifts tend to become more pronounced as $\alpha$ is increased.

\begin{figure*}[t!]
	\centering
	\includegraphics[width=0.9\textwidth]{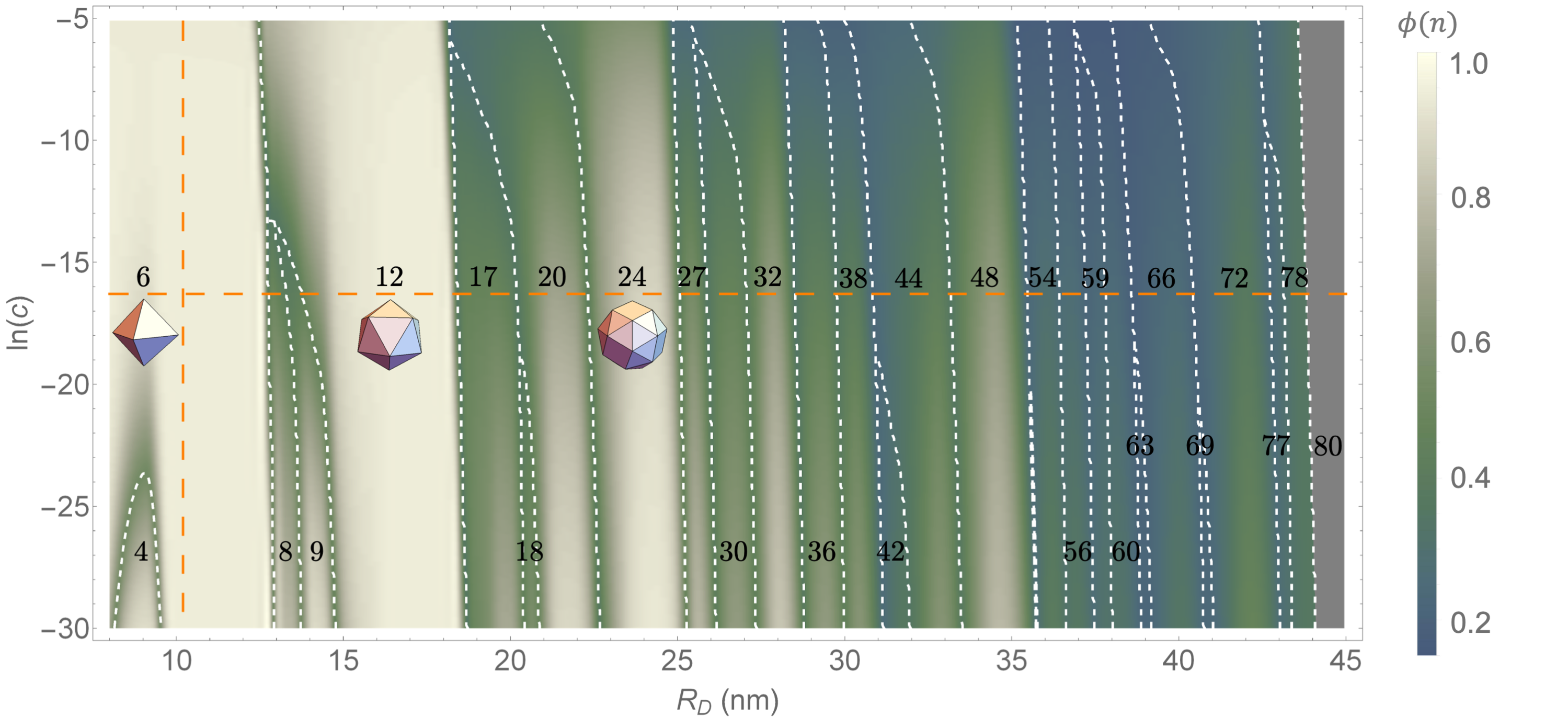} 
	\caption{MPPN self-assembly diagram for MPPNs formed from Piezo ion channels, obtained from the arclength parameterization of Eq.~(\ref{eqdefG}), as a function of the radius of curvature of the Piezo dome, $R_D$, and the protein number fraction in solution, $c$. Colors indicate the maximum $\phi(n)$ among all MPPN $n$-states considered. The color bar employed here is identical to the color bar employed in Fig.~\ref{Fig2}. As in Fig.~\ref{Fig2}, the dominant $n$-states of MPPNs are indicated by integers, and the white dashed curves show transitions in the dominant MPPN $n$-states. The horizontal dashed line indicates the protein number fraction $c\approx 7.8\times10^{-8}$ used in experiments on MPPNs formed from MscS \cite{Wu2013,basta14}, while the vertical dashed line shows the Piezo dome radius of curvature $R_D \approx 10.2$~nm observed for a closed state of Piezo \cite{Guo2017,saotome18,zhao18,wang19}. The polyhedra models illustrate dominant MPPN symmetries predicted by the MPPN self-assembly diagram, with $n=6$ (octahedron), $n=12$ (icosahedron), and $n=24$ (snub cube), respectively. Gray shading indicates the region of the MPPN self-assembly diagram dominated by MPPN $n$-states with $n=80$, which may be a spurious result of our constraint $3 \leq n \leq 80$.
	}
	\label{Fig4}
\end{figure*}

As described above, the self-assembly diagram in Fig.~\ref{Fig2}(a) was obtained by interpolating $\phi(n)$ between a discrete set of values of $\alpha$. Figure~\ref{Fig3}(a) compares the interpolated $\phi$ at the resolution $\Delta \alpha = 0.01$~rad used for Fig.~\ref{Fig2}(a) for selected (dominant) MPPN $n$-states at the protein number fraction $c \approx  7.8\times10^{-8}$ \cite{Wu2013,basta14} with the corresponding results of calculations done at a finer resolution $\Delta \alpha = 0.002$~rad. Figure~\ref{Fig3}(a) suggests that the interpolation scheme employed here provides accurate estimates of the dominant $\phi(n)$ for continuous $\alpha$. Figure~\ref{Fig3}(b) compares the results for $\phi(n)$ in Fig.~\ref{Fig2}(a), obtained with $A=A_D$ in Eq.~(\ref{AF}), with the corresponding results obtained with $A=A_S$. As expected, $A=A_S$ provides a good approximation for the weak membrane curvatures induced by MscS. But $A=A_S$ is, in general, not expected to give accurate results for membrane proteins such as Piezo that strongly curve the membrane. In the following we therefore focus on $A=A_D$ in Eq.~(\ref{AF}).

\begin{figure}[t!]
	\centering
	\includegraphics[width=\columnwidth]{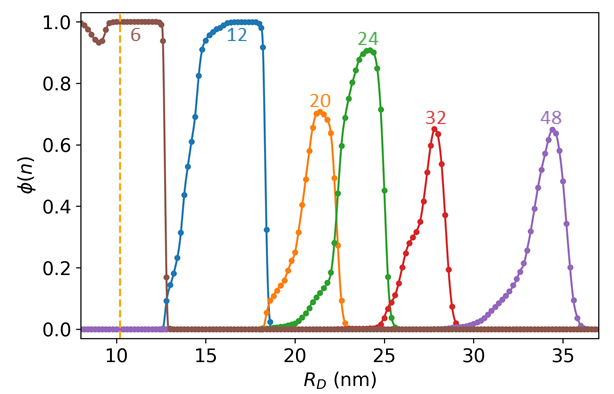} 
	\caption{Fractional abundance of MPPN $n$-states obtained from the arclength parameterization of Eq.~(\ref{eqdefG}), $\phi(n)$, versus radius of curvature of the Piezo dome, $R_D$, for selected (dominant) MPPN $n$-states in Fig.~\ref{Fig4} at the protein number fraction $c\approx 7.8\times10^{-8}$ used in experiments on MPPNs formed from MscS \cite{Wu2013,basta14}. As in Fig.~\ref{Fig4}, the vertical dashed line shows the Piezo dome radius of curvature $R_D \approx 10.2$~nm observed for a closed state of Piezo \cite{Guo2017,saotome18,zhao18,wang19}. All curves were obtained through interpolation of $\phi(n)$ with respect to $R_D$ for numerical results at a resolution $\Delta R_D = 0.2$~nm (dots), using third-order splines.
	}
	\label{Fig5}
\end{figure}

\textit{MPPN self-assembly from Piezo.} Figure~\ref{Fig4} shows the self-assembly diagram for MPPNs formed from Piezo ion channels as a function of the radius of curvature of the Piezo dome and the protein number fraction in solution. We obtained the results in Fig.~\ref{Fig4} using the  arclength parameterization of Eq.~(\ref{eqdefG}) with $A=A_D$ in Eq.~(\ref{AF}). The MPPN self-assembly diagram in Fig.~\ref{Fig4} includes highly curved MPPN states with, for instance, a contact angle $\alpha \approx 1.3$~rad at $R_D \approx 9.0$~nm. The vertical dashed line in Fig.~\ref{Fig4} indicates the Piezo dome radius of curvature $R_D \approx 10.2$~nm observed for a closed state of Piezo \cite{Guo2017,saotome18,zhao18,wang19} while the horizontal dashed line indicates, for reference, the protein number fraction $c \approx  7.8\times10^{-8}$ used in experiments on MPPNs formed from MscS \cite{Wu2013,basta14}. Figure~\ref{Fig5} shows, for this protein number fraction, the $\phi(n)$ associated with dominant MPPN $n$-states in Fig.~\ref{Fig4} as a function $R_D$. The results in Figs.~\ref{Fig4} and~\ref{Fig5} were obtained through (third-order spline) interpolation of $\phi(n)$ with respect to $R_D$, for numerical results at a resolution $\Delta R_D = 0.2$~nm. Similar results are obtained with a finer resolution in $R_D$.

We find in Fig.~\ref{Fig4} that the dominant MPPN $n$-state depends only weakly on the protein number fraction in solution, but strongly on the Piezo dome radius of curvature. As $R_D$ is increased, Piezo's membrane footprint becomes less curved \cite{Haselwandter2018}, yielding larger and less curved MPPNs that incorporate more Piezo proteins. Most notably, the MPPN self-assembly diagram in Fig.~\ref{Fig4} is dominated by highly symmetric MPPN $n$-states with octahedral ($O_h$; $n=6$), icosahedral ($I_h$; $n=12$), or snub cube ($O$; $n=24$) symmetry \cite{Clare1986,Clare1991}. For the protein number fraction $c \approx  7.8\times10^{-8}$ used in experiments on MPPNs formed from MscS \cite{Wu2013,basta14}, MPPN octahedra are dominant in Fig.~\ref{Fig4} for $8.0~\mathrm{nm} \lessapprox R_D \lessapprox 13$~nm, MPPN icosahedra are dominant for $14~\mathrm{nm} \lessapprox R_D \lessapprox 18$~nm, and MPPN snub cubes are dominant for $22~\mathrm{nm} \lessapprox R_D \lessapprox 25$~nm. At $c \approx  7.8\times10^{-8}$, the dominant MPPN states with $n=6$, 12, and 24 in Fig.~\ref{Fig4} have the characteristic MPPN radii $14~\mathrm{nm} \lessapprox R \lessapprox 15$~nm, $21~\mathrm{nm} \lessapprox R \lessapprox 25$~nm, and $30~\mathrm{nm} \lessapprox R \lessapprox 33$~nm, with the top of the Piezo dome being located approximately $1.3$--$9.2$~nm, $0.56$--$3.3$~nm, and $0.54$--$1.1$~nm above the spherical surface defined by $R$, respectively (Fig.~\ref{Fig1}).

Figures~\ref{Fig4} and~\ref{Fig5} show that, in addition to $n=6$, 12, and 24, MPPNs with $n=20$, 32, and 48 can also be abundant at $c \approx  7.8\times10^{-8}$. These MPPN $n$-states have $D_{3h}$-symmetry ($n=20$), $D_3$-symmetry ($n=32$), and $O$-symmetry ($n=48$)  \cite{Clare1986,Clare1991}. For $c \approx  7.8\times10^{-8}$, MPPN states with $n=20$, 32, and 48 are dominant for $20~\mathrm{nm} \lessapprox R_D \lessapprox 22$~nm, $27~\mathrm{nm} \lessapprox R_D \lessapprox 29$~nm, and $33~\mathrm{nm} \lessapprox R_D \lessapprox 36$~nm, and have the characteristic MPPN radii $28~\mathrm{nm} \lessapprox R \lessapprox 30~\mathrm{nm}$, $35~\mathrm{nm} \lessapprox R \lessapprox 36~\mathrm{nm}$, and $43~\mathrm{nm} \lessapprox R \lessapprox 44~\mathrm{nm}$, with the top of the Piezo dome being located approximately $0.73$--$1.35$~nm, $0.53$--$0.69$~nm, and $0.37$--$0.58$~nm above the spherical surface defined by $R$, respectively (Fig.~\ref{Fig1}). Overall, Figs.~\ref{Fig4} and~\ref{Fig5} thus predict that self-assembly of MPPNs from Piezo can yield highly symmetric and highly curved MPPN states, with the radius of curvature of the Piezo dome providing the critical control parameter for the symmetry and size of MPPNs formed from Piezo.

\textit{Conclusion.} We have developed here a methodology for predicting the symmetry and size of MPPNs with arbitrarily large (nonlinear) membrane curvature deformations. For MPPNs formed from MscS \cite{Bass2002,Steinbacher2007} this methodology predicts, with no adjustable parameters, the observed symmetry and size of MPPNs \cite{Wu2013,basta14}. Since MscS proteins only weakly curve the membrane \cite{Bass2002,Steinbacher2007,Phillips2009}, similar conclusions were reached previously using a small-gradient approximation \cite{li16,li17}. In contrast, (closed-state) Piezo proteins have a highly curved structure \cite{Guo2017,saotome18,zhao18,wang19}, and the resulting membrane shape deformations are highly nonlinear \cite{Haselwandter2018}. We find here that the self-assembly diagram for MPPNs formed from Piezo critically depends on the Piezo dome radius of curvature. In particular, for the Piezo dome radius of curvature $R_D \approx 10.2$~nm observed for a closed state of Piezo \cite{Guo2017,saotome18,zhao18,wang19}, we generally find MPPNs with six Piezo proteins and octahedral symmetry to be dominant. As the value of $R_D$ is increased, we find dominant MPPN states with icosahedral and snub cube symmetry, composed of 12 and 24 Piezo proteins, respectively. Such highly symmetric MPPN states may allow structural studies of MPPNs formed from Piezo \cite{zhang03,liu04,cockburn04} in the presence of transmembrane gradients similar to those found in cellular environments \cite{Wu2013,basta14}. Intriguingly, if gating of Piezo is accompanied by an increase in $R_D$ \cite{Guo2017,saotome18,zhao18,wang19,lin19,Haselwandter2018}, the distinct MPPN symmetries predicted here may be associated with distinct, biologically relevant conformational states of Piezo.

\textit{Acknowledgments.} We thank J. Agudo-Canalejo, O. Kahraman, D. Li, R. MacKinnon, and M. H. B. Stowell for useful discussions. This work was supported by NSF award number DMR-1554716 and the USC Center for High-Performance Computing.

\bibliographystyle{apsrev4-1}
\bibliography{Refs}

\end{document}